\documentclass[aps,prb,twocolumn,superscriptaddress,nofootinbib,preprintnumbers]{revtex4}

\usepackage{graphicx}
\usepackage{graphics}
\usepackage{amsmath,amssymb}
\usepackage[usenames]{color}
\usepackage{subfigure}
\usepackage{hyperref}
\usepackage{breakurl}
\usepackage{enumitem}
\usepackage{lgreek}

\newcommand{\be}{\begin{equation}}
\newcommand{\ee}{\end{equation}}
\newcommand{\bea}{\begin{eqnarray}}
\newcommand{\eea}{\end{eqnarray}}
\newcommand{\ben}{\begin{enumerate}}
\newcommand{\een}{\end{enumerate}}
\newcommand{\bit}{\begin{itemize}}
\newcommand{\eit}{\end{itemize}}

\newcommand{\la}[1]{\label{#1}}

\newcommand{\Eq}[1]{Eq.~(\ref{#1})}

\newcommand{\Sec}[1]{Sec.~\ref{#1}}

\newcommand{\Fig}[1]{Fig.~\ref{#1}}

\def\nl{\nonumber \\}

\newcommand{\vv}[1]{\mathbf #1}							% 3-vector
\newcommand{\bert}{\raise-0.45mm\hbox{\Large$\Box$}}			% D'Alembertian

\definecolor{BrickRed}{cmyk}{0,0.89,0.94,0.28}					%%%PANTONE 1805
\definecolor{MidnightBlue}{cmyk}{0.98,0.13,0,0.43}				%%%PANTONE 302
\definecolor{DarkGreen}{rgb}{0.100806,0.495968,0.209979}
\definecolor{orange}{rgb}{0.587167,0.354498,0.146197}

\begin{document}
	
\title{Dynamical theory for the battery's electromotive force}
	
\author{Robert Alicki}
\email{robert.alicki@ug.edu.pl}
\affiliation{International Centre for Theory of Quantum Technologies (ICTQT), University of Gda\'nsk, 80-308, Gda\'nsk, Poland}
\author{David Gelbwaser-Klimovsky}
\email{dgelbi@mit.edu}
\affiliation{Physics of Living Systems, Department of Physics, Massachusetts Institute of Technology, Cambridge, MA 02139, USA}
\author{Alejandro Jenkins}
\email{alejandro.jenkins@ucr.ac.cr}
\affiliation{International Centre for Theory of Quantum Technologies (ICTQT), University of Gda\'nsk, 80-308, Gda\'nsk, Poland}
\affiliation{Laboratorio de F\'isica Te\'orica y Computacional, Escuela de F\'isica, Universidad de Costa Rica, 11501-2060, San Jos\'e, Costa Rica}
\author{Elizabeth von Hauff}
\email{e.l.von.hauff@vu.nl}
\affiliation{Department of Physics and Astronomy, Vrije Universiteit Amsterdam, De Boelelaan 1081, 1081 HV Amsterdam, The Netherlands}
	
\date{First version, 30 Oct.\ 2020; last revision, 25 Mar.\ 2021.  To be published in Phys.\ Chem.\ Chem.\ Phys.}

\begin{abstract}
We propose a dynamical theory of how the chemical energy stored in a battery generates the electromotive force (emf).  In this picture, the battery's half-cell acts as an engine, cyclically extracting work from its underlying chemical disequilibrium.  We show that the double layer at the electrode-electrolyte interface can exhibit a rapid self-oscillation that pumps an electric current, thus accounting for the persistent conversion of chemical energy into electrical work equal to the emf times the separated charge.  We suggest a connection between this mechanism and the slow self-oscillations observed in various electrochemical cells, including batteries, as well as the enhancement of the current observed when ultrasound is applied to the half-cell.  Finally, we propose more direct experimental tests of the predictions of this dynamical theory.
\end{abstract}
	
\maketitle

%\tableofcontents

%%%%%%%%%%
%%% INTRODUCTION
%%%%%%%%%%

\section{Introduction}
\la{sec:intro}

Macroscopic devices capable of delivering sustained power, such as motors, turbines, generators, and combustion engines, have moving parts whose cyclical dynamics govern the transformation of potential energy from an external source into useful work.\cite{SO}  In contrast, the operation of solid-state and electrochemical devices such as photovoltaic cells, thermoelectric generators, and batteries, relies on an electromotive force (emf) in order to convert chemical potential into electrical power. 

Volta introduced the concept of emf in 1798 to refer to the mechanism that causes and maintains the separation of opposite charges in a battery, despite the electrostatic attraction between them.\cite{emf}  Volta's discoveries triggered intense debates on the seat and mechanism of the battery's emf.  These involved the leading physicists and chemists of the 19th and early 20th centuries, including Biot, Davy, Ohm, Berzelius, A.~C.~Becquerel, Faraday, Helmholtz, Kelvin, Maxwell, Heaviside, Lodge, Ostwald, Nernst, and Langmuir.\cite{Kragh}  Rather that reaching a decisive resolution, this scientific controversy petered out with the practical advances in electrochemistry and the shift in focus resulting from the quantum revolution.\cite{Chang1, Chang2}

We know now that the battery's emf arises because a chemical reaction ``yields more energy than it costs to buck the [general electric] field.''\cite{Purcell} However, even this correct observation fails to provide a realistic dynamical account of the origin and nature of the emf responsible for the power output of the device during discharge.\cite{emf, Saslow}  We believe that this is because the theoretical treatment of electrochemical cells is still largely based on electrostatics.  Such a description can account for the flow of current through the external circuit (from high to low electrostatic potential), as in the case of a discharging capacitor.  But, though it provides the correct energy budget for the conversion of chemical into electrical energy, it cannot describe the dynamics responsible for the flow of current {\it within} the battery that generates and maintains the electric potential difference between the terminals.

Physicist Peter Heller suggested replacing the term emf by {\it electromotive pump} (emp), to describe any underlying physical mechanism that promotes the circulation of electric current around a closed path.\cite{Saslow}  It is well understood that the battery's emf results from the action of ``surface pumps'' at the electrode-electrolyte interfaces within the device, and that the energy consumed by this pumping comes from the redox chemical reactions at those interfaces.\cite{Saslow, Baierlein}  However the often cited explanation of this pumping in terms of an ``equivalent force'' in diffusion theory has been characterized as merely ``heuristic''.\cite{emf}  Describing the pumping of charge at the electrode-electrolyte interface using the equations for diffusion with fixed-concentration boundary conditions presents serious difficulties due to the indefiniteness of the velocity of Brownian particles.\cite{Singer} Ultimately, we are still lacking a clear quantitative description of the microphysics at the electrode-electrolyte interface.\cite{Saslow} 

In this article we propose that the pumping of charge that generates the emf of a battery is associated with dynamics-in-time of the double layer at the electrode-electrolyte interface. This double layer incorporates a mechanical elasticity\cite{EC} or ``squishiness'',\cite{squishy1, squishy2} that is compatible with an electrochemical Gouy--Chapman-type model.  The mutual coupling between the mechanical and electrical degrees of freedom can cause a dynamical instability leading to a self-oscillation.  The self-oscillating double layer acts as an internal piston for the half-cell, thereby pumping current against the average electrostatic potential.\cite{LEC}

In our picture, the battery's emf is analogous to the ``head rise'' produced by a hydraulic pump.  In the mechanical engineering literature, the head rise times the flow rate is equated to the external power consumed by the pump, minus the pump's internal dissipative losses.  In an ideal hydrodynamic pump (much like in a battery), the pressure is nearly independent of the flow, up to some maximum flow rate beyond which the pump cannot run properly.\cite{head}

We predict that the pumping of charge within the battery must be associated with high-frequency oscillations that actively pump current at the double layer.  Much slower self-oscillations have been observed in various electrochemical systems,\cite{koper} including modified Li-ion batteries.\cite{Li}  We interpret those slow oscillations as secondary effects of the battery's pumping dynamics, effects that may become evident towards the end of the discharging process as the battery's operation becomes diffusion-limited and the steady pumping regime is destabilized.  Similar slow self-oscillations have also been reported in the catalytic reactions\cite{Ertl, Delmonde} and in self-charging electrochemical cells with a ferroelectric glass electrolyte.\cite{Goodenough}

Some authors have proposed exploiting the mechanical energy of electrochemical self-oscillations to perform useful work,\cite{Isakova} but the experimentally observed electrochemical oscillations have not previously been connected to the pumping dynamic responsible for the battery's emf.  Analogous self-oscillations, with frequencies well below that of the pump's running, are often seen in hydraulic pumps (where they can pose serious problems for the device's proper functioning).\cite{hydro-so}

%%%%%%%%%%
%%% SUPERCAPACITORS VS. BATTERIES
%%%%%%%%%%

\section{Capacitors vs.\ batteries}
\la{sec:supercapacitors}

Like a battery, a ``supercapacitor'' has an electrochemical double layer at each of the two electrode-electrolyte interfaces.\cite{supercaps}    The formation of the double layer is a spontaneous process that can be understood in terms of an equilibration of the electrochemical potential.\cite{doublelayer}  The gain in chemical energy (which is linear in the number of separated charges) is balanced by the electrostatic energy accumulated (which is quadratic).  The presence of this double layer ---with the corresponding difference in the electrostatic potential between the electrode and the electrolyte--- does not by itself imply an emf.\cite{Roberts}  Indeed, comparison of the discharge curves for the supercapacitor and the battery reveals a clear qualitative difference between the two devices.

\begin{figure}[t]
\centering
	\subfigure[]{\includegraphics[height=0.16 \textwidth]{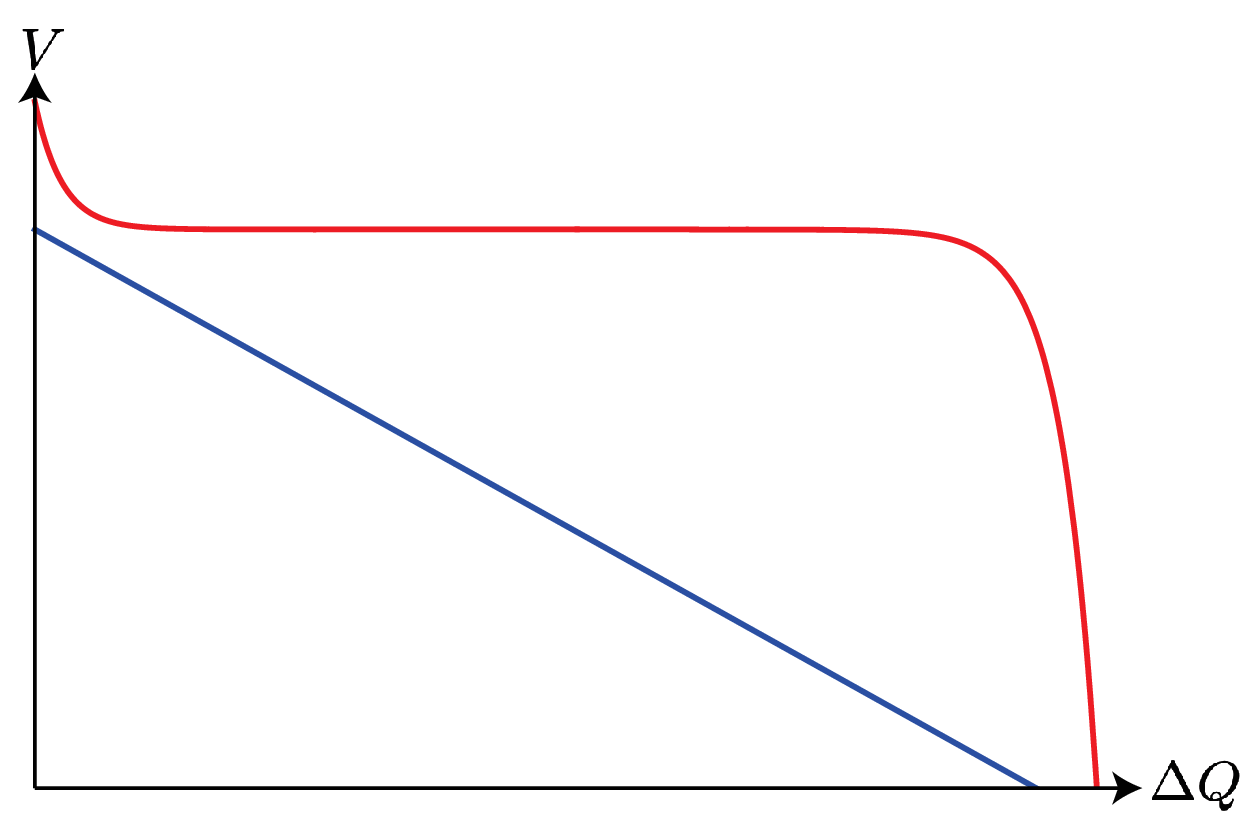}}
	\subfigure[]{\includegraphics[height=0.21 \textwidth]{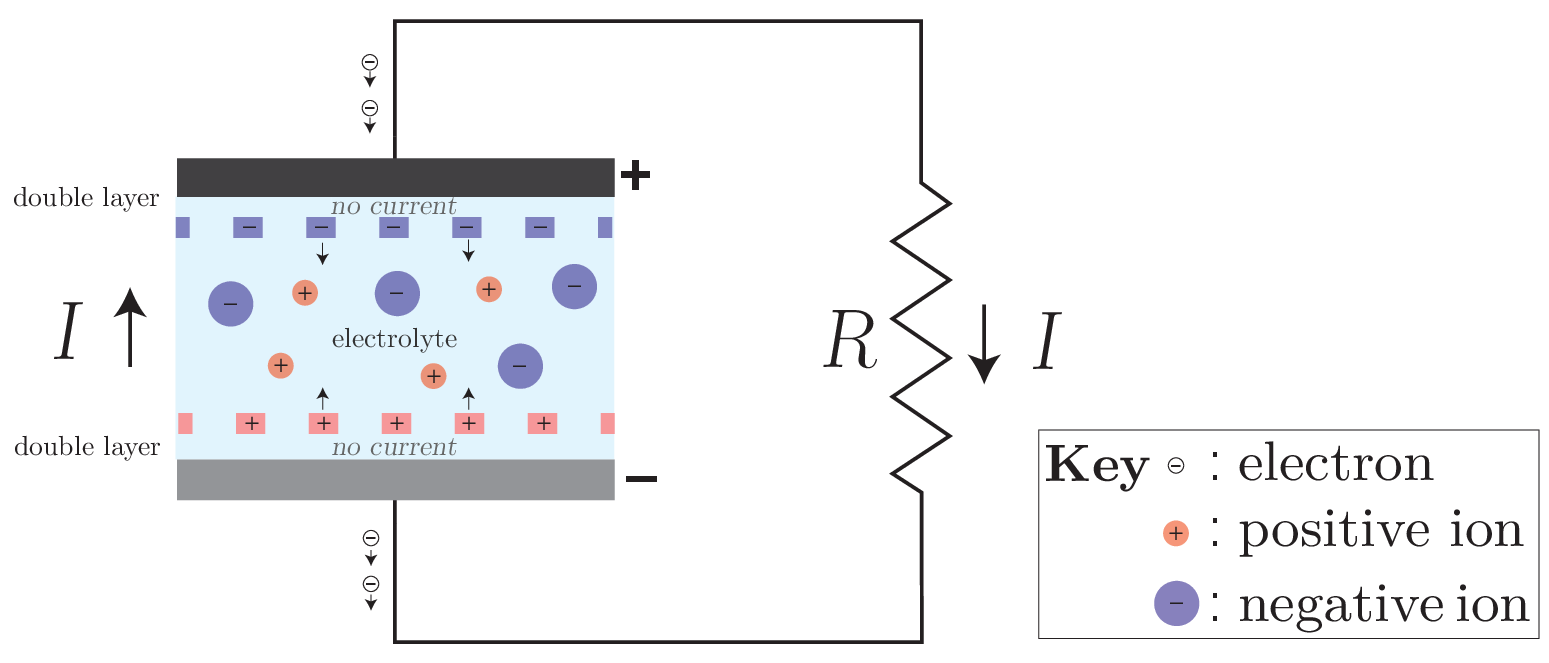}}
	\subfigure[]{\includegraphics[height=0.21 \textwidth]{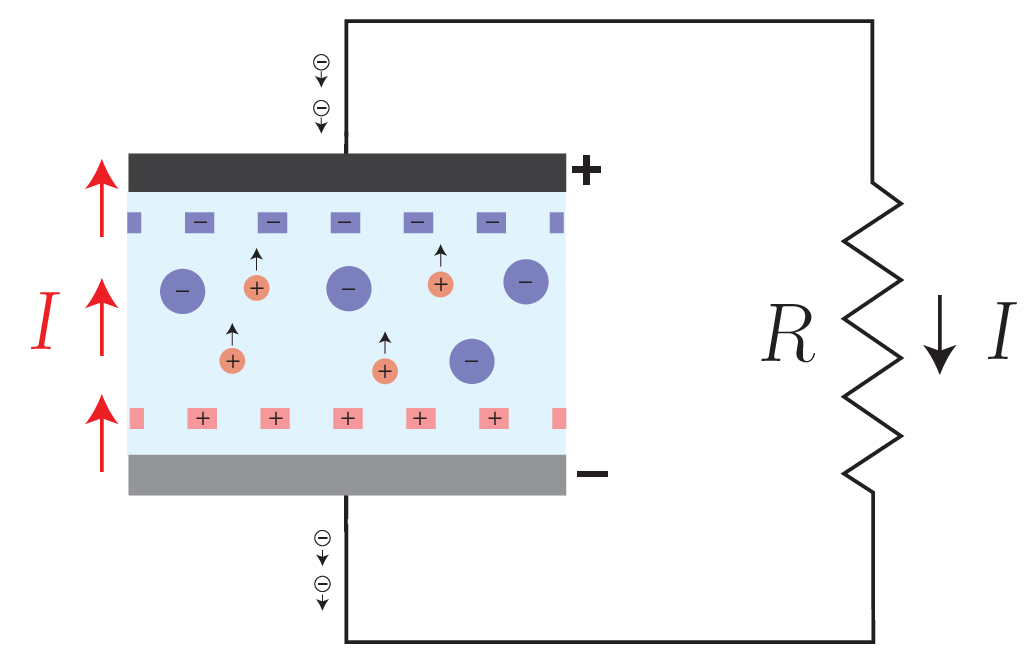}}
\caption{\small Image (a) shows the voltage $V$ versus integrated current $\Delta Q \equiv \int_0^t I(t') dt'$ of an ideal supercapacitor (blue curve) and battery (red curve).  The other images schematically show the electrical current $I$ in a circuit connected to: (b) a supercapacitor and (c) a battery.  The voltage $V$ is the electrostatic potential difference between the $\boldsymbol +$ and $\boldsymbol -$ terminals of each device.\label{fig:cap-batt}}
\end{figure}

Figure \ref{fig:cap-batt}(a) shows a capacitor's linear relation between voltage and integrated current (blue curve), which can be entirely understood in terms of electrostatics. If the capacitor is charged, and its terminals are subsequently connected to an external load, a current flows between the terminals at the expense of the discharging potential $V$. The capacitor cannot drive current along a closed path,  and therefore has no emf. A rechargeable electrochemical battery, on the other hand has a non-linear (and, therefore, non-electrostatic) relation between voltage and integrated current, shown by the red curve in Fig. \ref{fig:cap-batt}(a).  This reflects the presence of an emf within the battery, which keeps the voltage difference between the terminals nearly constant until the chemical reservoir becomes depleted.\cite{supercaps}

The charging of a supercapacitor involves only reversible separation of charges, i.e., the storage of electrostatic potential energy. We will argue that the charging of a battery, on the other hand, produces a ``chemical fuel'', which is then irreversibly consumed during discharge.  The battery's half-cell acts like a {\it chemical engine}, powering a charge pump at the electrode-electrolyte interface that drives the circulation of current in the external circuit.  The average ``pressure'' produced by the ``pump's piston'' acting on the electron fluid is the source of the battery's emf, which remains almost constant up to the moment when the chemical fuel runs out.

The emf in the battery is often equated to the voltage measured at the terminals, but this is a conceptual error, as the authors of Ref.~\onlinecite{emf} underline.  In open-circuit conditions, the potential $V_{\rm oc}$ is indeed equal to the emf $\cal E$, and the relation
\be
V_{\rm oc} = {\cal E}
\la{ec:Voc}
\ee
provides an accurate {\it measurement} of the emf.  However, even the zero-current limit of the battery's operation cannot be understood electrostatically, and the relation between emf and voltage at the terminals becomes more involved as the battery is operated away from open-circuit conditions.  The mathematics of the emf as a form of pumping is discussed in detail in Ref.~\onlinecite{LEC}.  In \Sec{sec:double-layer} we propose a model for the physical origin of the non-electrostatic force that gives rise to the battery's emf.

As shown in \Fig{fig:cap-batt}(b), the current that a charged supercapacitor generates when connected to an external load $R$ is not closed: there is a current $I$ both through the load (given by a flow of electrons) and through the bulk of the electrolyte (given by diffusion of ions), but no current flows through the double layer at each electrode-electrolyte interface.  This is equivalent to replacing the supercapacitor by two simple capacitors (each corresponding to one of the double layers) connected in series.  On the other hand, as illustrated in \Fig{fig:cap-batt}(c), for the battery the circuit is closed by a ballistic flow of ions within the electrolyte and {\it through the double layers}.
	
Electronics textbooks distinguish between {\it active} devices that can amplify the power that they receive from the circuit, and {\it passive} devices that cannot.  In this classification, the supercapacitor, like the ordinary capacitor, is passive.  Horowitz and Hill note that active devices ``are distinguishable by their ability to make oscillators, by feeding from output signal back into the input,'' i.e., to self-oscillate.\cite{HH}  The self-oscillations reported in Refs.~\onlinecite{Li} and \onlinecite{Goodenough} can therefore be interpreted as evidence that the battery (considered as a circuit element, rather than as just an external power source) is active.

The distinction between active and passive devices offers an instructive way of framing the key qualitative distinction between the battery and the supercapacitor.  A passive device can consume free energy, but it cannot use it to perform {\it sustained} work or to pump a flow.  An active system, on the other hand, uses some of the free energy that it consumes from an external source in order to generate an active, non-conservative force, which can be used to pump a flow against an external potential or to sustain a circulation.\cite{LEC}  The emf corresponds to that non-conservative force per unit charge, integrated over the closed path of the current.  We return to this specific issue in \Sec{sec:pump}.

%%%%%%%%%%
%%% DOUBLE-LAYER DYNAMICS
%%%%%%%%%%

\section{Active double-layer dynamics}
\la{sec:double-layer}

\begin{figure} [t]
	\begin{center}
		\includegraphics[width=0.13 \textwidth]{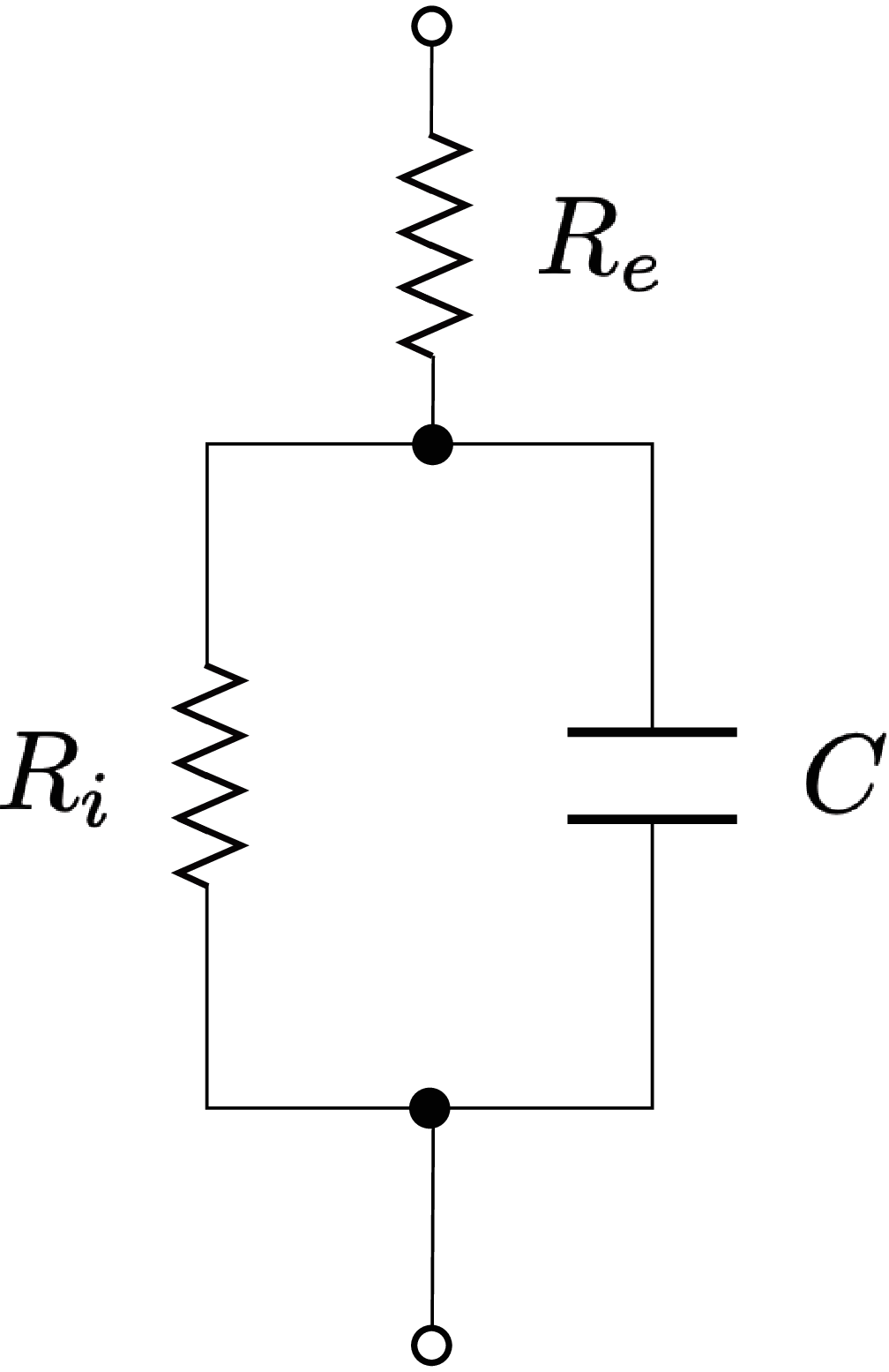}
	\end{center}
\caption{\small Equivalent AC circuit for the half-cell, in the case in which the impedance is dominated by the electron transfer resistance: $R_e$ is the electrolyte resistance, $C$ is the double layer's capacitance, and $R_i$ is the double layer's internal resistance.  See Refs.~\onlinecite{Hamann, impedance}.\label{fig:equivalent}}
\end{figure}

Macroscopic engines generate an active, non-conservative force, via thermodynamically irreversible processes involving a positive feedback.  This allows them to do work persistently (i.e., cyclically), at the expense of the external disequilibrium.  In many cases this work generation appears as the self-oscillation of a piston.\cite{SO, LeCorbeiller1, LeCorbeiller2}  The macroscopic kinetic energy of this self-oscillation can be used to pump a flow against an external potential, or along a closed path.

More specifically, a self-oscillator is a system that can generate and maintain a periodic motion at the expense of a power source with no corresponding periodicity.\cite{Andronov}  Such a process is necessarily dissipative and requires a positive feedback between the oscillating system and the action upon it of the external power source.\cite{SO}  Here we will describe the pumping of charge within a battery in terms of the self-oscillation of the electrochemical double layer of the half-cell.  This is a direct application to a Gouy--Chapman-type model of the ``leaking elastic capacitor'' (LEC) model that has recently been worked out mathematically in Ref. \onlinecite{LEC}. That work, in turn, is closely related to the theory of the ``electron shuttle'' as an autonomous engine powered by an external disequilibrium in chemical potential.\cite{gorelik, wachtlerNJP, wachtlerPRA}

The passive half-cell can be described by the AC equivalent circuit model shown in \Fig{fig:equivalent}.\cite{Hamann, impedance}  The capacitance $C$ of the double layer is not fixed, but rather increases as the potential difference between the Helmholtz plane of ions and the electrode surface is increased.  In the Gouy--Chapman model this is explained as the result of the diffuse layer of ions in the electrolyte becoming more compact as the applied potential is increased.  It is well established that such a re-arrangement of charges at the solid-liquid interface can lead to dynamical instabilities.\cite{squishy1, squishy2}  In this section we shall extend this into a dynamical description that can explain the pumping of charge that generates the battery's emf.

\subsection{Leaking elastic capacitor model}
\la{sec:LEC}

\begin{figure} [t]
	\centering
	\includegraphics[width=0.35 \textwidth]{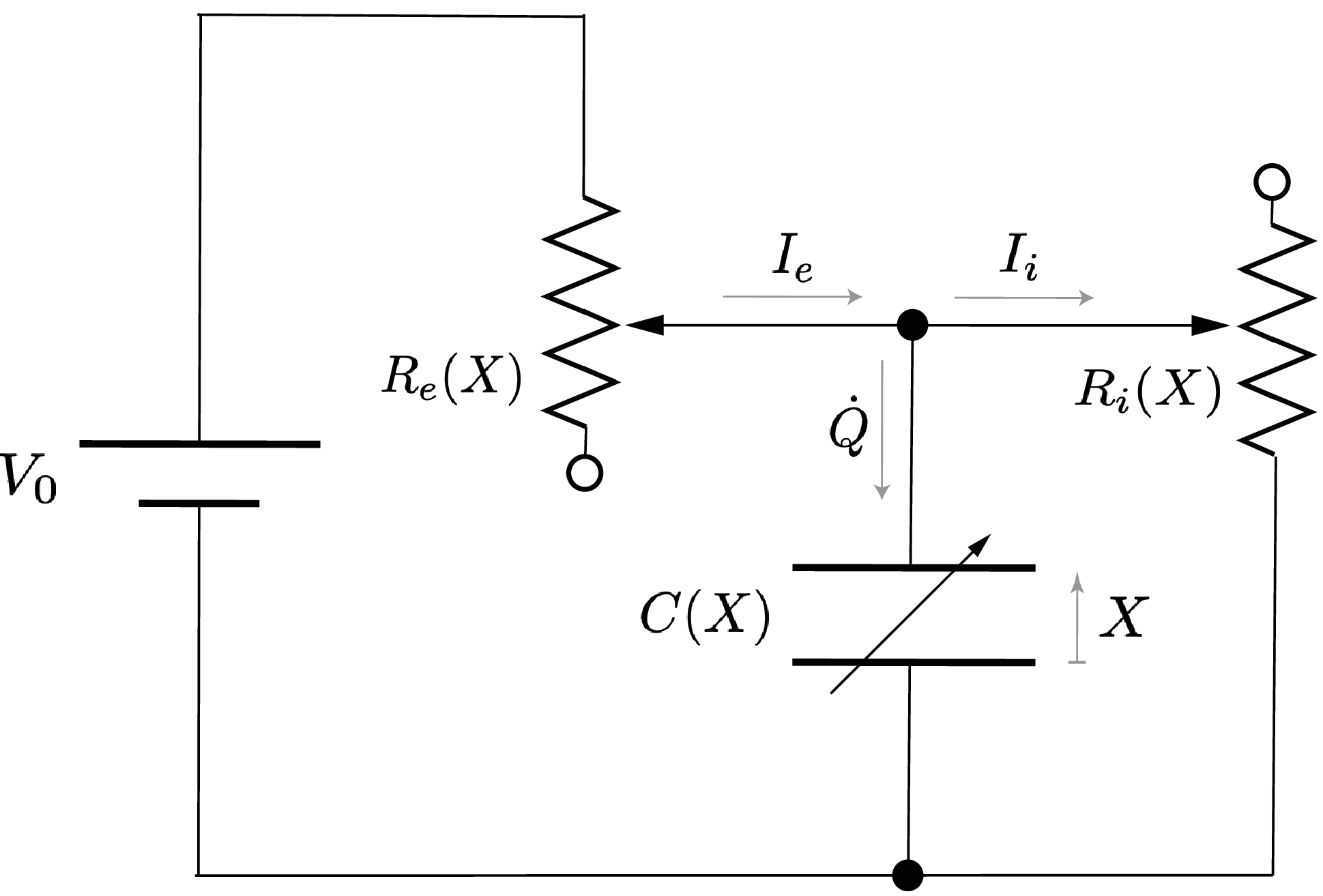}
\caption{\small Equivalent circuit for the leaking elastic capacitor (LEC).  Adapted from Ref.~\onlinecite{LEC}.\label{fig:LEC}}
\end{figure}

In order to describe the dynamics of the {\it active} half-cell in a battery, we consider the general mathematical description of the LEC model introduced in Ref.~\onlinecite{LEC}. Figure \ref{fig:LEC} shows the equivalent electromechanical circuit for the LEC, consisting of the external source of energy (the voltage $V_0$), a capacitance $C$, an external resistance $R_e$ in series with $C$, and internal resistance $R_i$ in parallel with $C$. The capacitance $C$, as well as the resistances $R_{e, i}$, are presumed to be a functions of the distance $X$ between the capacitor plates.  Here $V_0$ represents a simple electrostatic potential.

Let $Q$ be the total charge accumulated by the capacitor.  According to Kirchhoff's rules, the current $I_e$ through resistor $R_e$ is equal to
\be
I_e = \dot Q + I_i ,
\la{eq:Kirch1}
\ee
where $I_i$ is the current through resistor $R_i$, and
\be
V_0 = I_e R_e + I_i R_i .
\la{eq:Kirch2}
\ee
The voltage drop across the capacitor is
\be
Q / C = I_i R_i .
\la{eq:Kirch3}
\ee
Combining Eqs.\ \eqref{eq:Kirch1}, \eqref{eq:Kirch2}, and \eqref{eq:Kirch3} yields a dynamical equation for the total charge $Q(t)$ stored in the LEC of \Fig{fig:LEC}:
	\be
\dot Q =  - \left[ \frac{1}{C(X) R_e (X)} + \frac{1}{C(X) R_i (X)} \right] Q + \frac{V_0}{ R_e (X)} .
\la{eq:Kirch}
\ee
Assuming that one of the plates of the capacitor is fixed while the other is free to move, Newton's second law of motion yields a dynamical equation for $X(t)$ of the form
\be
\ddot X + \gamma \dot X = \frac 1 M f(X, Q) .
\la{eq:Newt}
\ee
In \Eq{eq:Newt} $\gamma$ is a viscous damping coefficient, $M$ is the mass of the moving capacitor plate, and the force $f(X, Q)$ is the sum of electrostatic attraction and molecular repulsion.  For simplicity and definiteness, we model the repulsion with a phenomenological parameter $\sigma > 0$:
\be
f(X, Q) = -\frac{Q^2}{2X C(X)} + \frac{\sigma}{X} .
\la{eq:f}
\ee
The choice of the repulsive force in \eqref{eq:f} as $\sim X^{-1}$ is motivated by a simple picture of the pressure exerted by a gas of ions confined within the double layer, as in the Gouy-Chapman model.\cite{Gouy} Note that it is consistent with the fact that if $Q \to 0$ then the equilibrium separation $X$ diverges, since in that case the Helmholtz layer of ions dissolves away.  The phenomenological parameter $\sigma$ in \Eq{eq:f} can be taken from the value of $X_0$ corresponding to the equilibrium width of the double layer.  Our conclusions are essentially independent of the details of this non-linear repulsive term.   For recent work on a detailed modeling of the mechanical properties of the electrochemical double layer, see Ref.~\onlinecite{Monroe}.

This simple electromechanical system can display a range of qualitatively different dynamical behaviors, including fixed-point solutions, regular limit cycles, and chaotic behavior; the details of this are discussed in Ref.~\onlinecite{LEC}.  The physical origins of self-oscillations in this model can be qualitatively understood without detailed mathematical analysis.  A change in $X$ implies changes to the internal and external resistances $R_{i,e}$.  These modulate the rate with which current $Q$ flows into or out of the capacitor.  If the charge $Q$ is modulated in phase with $- \dot X$, the electrostatic attraction between the two plates will increase when the LEC is contracting, and decrease when the LEC is expanding.  Such a positive feedback between $X$ and $Q$ will {\it anti-damp} the mechanical oscillation.  If this feedback-induced anti-damping exceeds the viscous damping $\gamma$, the equilibrium configuration becomes unstable and small oscillations about the equilibrium configuration grow exponentially in amplitude until they are limited by the non-linearity in \Eq{eq:f}.

The damping coefficient can be decomposed as
\be
\gamma = \gamma_{\rm diss} + \gamma_{\rm load} ,
\la{eq:gammas}
\ee
where $\gamma_{\rm diss}$ gives the approximately constant contribution from internal (viscous) dissipation and $\gamma_{\rm load}$ models energy loss of the oscillating double layer due to pumping of electric current in the external circuit (load).  This distinction will be discussed in detail in \Sec{sec:thermo}.

\subsection{Chemical engine}
\la{sec:chemical}

To describe the dynamics of the double layer within the battery we must replace \Eq{eq:Kirch}, which was based on a purely electromechanical model, by a suitable chemical kinetic equation.  For definiteness and simplicity, consider a redox reaction of the form
\be
AB  \rightleftharpoons B + A^+   + e^- ,
\la{eq:redox}
\ee
as in a Li-ion battery.  The forward reaction in \Eq{eq:redox} increases the absolute value of $Q$ by injecting electrons into the ``electrode plate'' and positive ions into the ``electrolyte plate'' (corresponding to the Helmholtz plane of ions). According to the law of mass action,
\be
\frac{d [e^-]}{dt} = k_+ \, [AB] - k_- \, [B] [A^+] [e^-] ,
\la{eq:mass-action}
\ee
where $[ \, \cdot \, ]$ represents the concentration of the corresponding species in \Eq{eq:redox}, within the effective volume near the electrode surface in which the reactions take place.  Meanwhile, $k_\pm$ are the ``rate constants'', which in this case depend on the state of the double layer. These rates are related to the Gibbs free energy of reaction, $\Delta G(X, Q)$, by the formula
	\be
	\ln K_{\rm eq}(X,Q) = \ln \frac {k_+ (X, Q)}{k_-(X,Q)} = -\frac{\Delta G(X, Q)}{RT} + c ,
	\la{eq:Gibbs}
	\ee
	where  $K_{\rm eq}$ is the equilibrium constant, $T$ is the temperature and $R$ the universal gas constant, and $c$ is a constant depending on the units of concentration.

We assume that the concentrations $[AB]$ and $[B]$ in the double layer are kept constant by diffusion in the electrolyte, while $[e^-]$ is dynamical.  The change in $[e^-]$ is accompanied by the corresponding variation of $[A^+]$, so as to maintain the net charge neutrality of the double layer.  But the average of $[A^+]$, inside the effective volume associated with the reaction of \Eq{eq:mass-action}, is much larger than the variation of $[e^-]$, because the double layer is sparse compared to the density of atoms and of ions in solution: an estimate for the Li-ion battery, using the parameters reported in Ref.~\onlinecite{Favaro2016}, gives a density $\sim 10^{-3} q /\hbox{\AA}^2$, where $q$ is the elementary charge.  We may therefore approximate $[A^+]$ as constant in \Eq{eq:mass-action}.

Stability of the equilibrium state for the double layer produced by the redox reaction implies that 
\be
\frac{\partial}{\partial X}\Delta G(X_0, Q_0) =\frac{\partial}{\partial Q}\Delta G(X_0, Q_0) = 0 .
\la{eq:Gibbs-stab}
\ee
In order to obtain the engine dynamics that we shall associate with the pumping of charge inside the battery, we must consider an additional process not contemplated in \Eq{eq:mass-action}:  A {\it leakage} of charge is needed to perturb the equilibrium state $(X_0 , Q_0)$ of the double layer and the corresponding potential difference $V_d(X_0, Q_0)$.  This leakage reduces $V_d$, inducing a chemical reaction that tends to restore equilibrium, in accordance with Le Ch\^atelier's principle. Under specific conditions that we will determine, the system then overshoots the equilibrium configuration and the double layer goes into a persistent oscillation.  We will show that such a self-oscillation can give the pumping necessary to account for the generation of the battery's emf.

The charge of the double layer is
\be
Q = [e^-] \cdot q {\cal A} \ell,
\ee
where $\cal A$ is the total surface area (an extensive parameter) and $\ell$ is the effective width of the region containing the double layer in which the reaction takes place (an intensive parameter).  Combining the chemical reaction described by \Eq{eq:mass-action} with the leakage of the double layer gives us a dynamical equation for $Q$ of a form very similar to \Eq{eq:Kirch}:
 \be
\dot Q = - r_- (X,Q) Q + q r_+(X,Q) ,
\la{eq:kinQ}
\ee
with
\bea
r_- (X,Q) &=& k_-(X, Q) [B][A^+] + k_{\rm out} , \nl
r_+ (X,Q) &=& k_+(X,Q) [AB] {\cal A} \ell .
\la{eq:rates}
\eea
In \Eq{eq:rates} the term $k_{\rm out}$ describes the leakage of charge from the double layer due to an internal resistance and to consumption of current by an external load connected to the battery's terminals.  This $k_{\rm out}$ is intensive (i.e., it does not depend on the size of the system, given by the area $\cal A$).

\subsection{Conditions for self-oscillation}
\la{sec:conditions}

Note that the rates $r_\pm$ in \Eq{eq:kinQ} depend on both $Q$ and $X$.  This is more general than \Eq{eq:Kirch} for the purely electromechanical system, but the basic mechanism of feedback-induced self-oscillation is qualitatively similar in both cases.

According to the calculations detailed in the Appendix, the necessary condition for self-oscillation of the electrochemical double layer (i.e., for positive feedback between $X$ and $Q$) is
\bea
0 < b &=& r_-(X_0,Q_0) X_0 \frac{\partial}{\partial X}\ln \left[ \frac{r_+(X, Q_0)}{r_-(X, Q_0)} \right]_{X = X_0} \nl
&=&k_{\rm out} \, X_0 \frac{\partial}{\partial X} \ln k_-(X_0,Q_0) .
\la{eq:b-cond}
\eea
Positivity of ${\partial_ X}\ln k_-$ follows from the fact that increasing $X$ increases the electrostatic potential difference $V_d (X, Q) = Q/C(X)$.  This favors the reverse reaction in \Eq{eq:redox}, in which the positive ion $A^+$ goes from the positively charged Helmholtz layer and into the negatively charged electrode, i.e., down in the potential difference $V_d$.  Larger $X$ therefore increases the reaction rate $k_-$ in \Eq{eq:redox}.

In particular, when 
\be
b > \frac 1 2 \gamma
\la{eq:bfactor}
\ee
the stationary solution becomes unstable and any small perturbation will give rise to a self-oscillation, which in the linear regime has angular frequency
\be
\Omega_0 = \sqrt{\epsilon\epsilon_0} \frac{V_d(X_0 , Q_0)}{\sqrt{2\rho }X_0^2}
\la{eq:Omega}
\ee
and exponentially growing amplitude.  Equation \eqref{eq:Omega} is expressed in terms of experimental parameters characterizing the electrochemical double layer, namely the equilibrium width of the double layer $X_0$, the potential drop $V_d (X, Q) =  Q / C(X)$ at $\{X_0, Q_0\}$ over the double layer, which is practically equal to the measured potential (i.e., the average electrostatic potential over a complete oscillation cycle), the density of the electrolyte $\rho$, and its permittivity $\epsilon \epsilon_0$.

The condition of \Eq{eq:bfactor} corresponds to the ``Hopf bifurcation'' of the dynamical system, at which the equilibrium becomes unstable due to the anti-damping of small oscillations.\cite{Strogatz} As the amplitude of such an oscillation increases, non-linearities become important so that eventually a limit-cycle regime is reached, giving a regular oscillation with steady amplitude.\cite{SO, Strogatz} 

\subsection{Thermodynamic interpretation}
\la{sec:thermo}

Note that, despite the mathematical similarity between Eqs.\ \eqref{eq:Kirch} and \eqref{eq:kinQ}, the physics that they describe is different.  In the electromechanical model of the LEC, based on \Eq{eq:Kirch}, all electric currents are driven by the external voltage source $V_0$, and the mechanical energy of the self-oscillation is simply dissipated.  On the other hand, in the case of the model for the battery's half-cell, based on \Eq{eq:kinQ}, electric current can leave the system and circulate in an external circuit, {\it pumped} by the mechanical self-oscillations of the double layer.  That is, part of the mechanical energy of the self-oscillation is dissipated and part of it is transformed into an electrical work $W = {\cal E} \, \bar Q$, where $\cal E$ is the emf and $\bar Q$ is the total charge driven around the closed circuit.

Let us first consider this process of transformation of chemical energy into an emf from the point of view of thermodynamics.  The total energy of the LEC can be expressed as 
\bea
E =  \frac 1 2 M \dot{X}^2 +  U(X, Q) , \nl
f(X,Q) = -\frac{\partial}{\partial X} U(X,Q) ,
\la{eq:E}
\eea
with time derivative
\be
\dot E =  -( \gamma_{\rm diss} +\gamma_{\rm load} )  M \dot X^2 +  V_d(Q,X) \dot Q .
\la{eq:dE}
\ee
Equation \eqref{eq:dE} can be interpreted in terms of the first law of thermodynamics.  Taking the system of interest to be the battery's half-cell, we have that
\be
dE = \delta {\cal Q} - \delta W + \mu \, d N ,
\la{eq:1law}
\ee
with internal energy $E$, heat flow to the environment \hbox{$-\dot{\cal Q} =  \gamma_{\rm diss} \, M \dot X^2$}, and electric power output \hbox{$\dot W = \gamma_{\rm load} \, M \dot X^2$}. We identify the electrochemical potential $\mu$ in \Eq{eq:1law} with $q V_d(Q,X)$, and the quantity of matter $N$ with $Q/q$.

It should be stressed that the two damping coefficients $\gamma_{\rm diss}$ and $\gamma_{\rm load}$, which appear in Eqs.\ \eqref{eq:gammas} and \eqref{eq:dE}, are qualitatively different from a thermodynamic point of view: $\gamma_{\rm diss}$ represents the dissipation of mechanical energy into the disordered motion of the microscopic components of the environment in which the double layer is immersed, while $\gamma_{\rm load}$ describes the transfer of energy from the self-oscillating double layer into the pumping of a coherent, macroscopic current that carries no entropy.  The fluctuation-dissipation theorem therefore applies only to $\gamma_{\rm diss}$, a point that should be born in mind if the presentmodel were extended to incorporate thermal noise via a Fokker-Planck equation.

\begin{figure} [t]
	\centering
	\includegraphics[width=0.2 \textwidth]{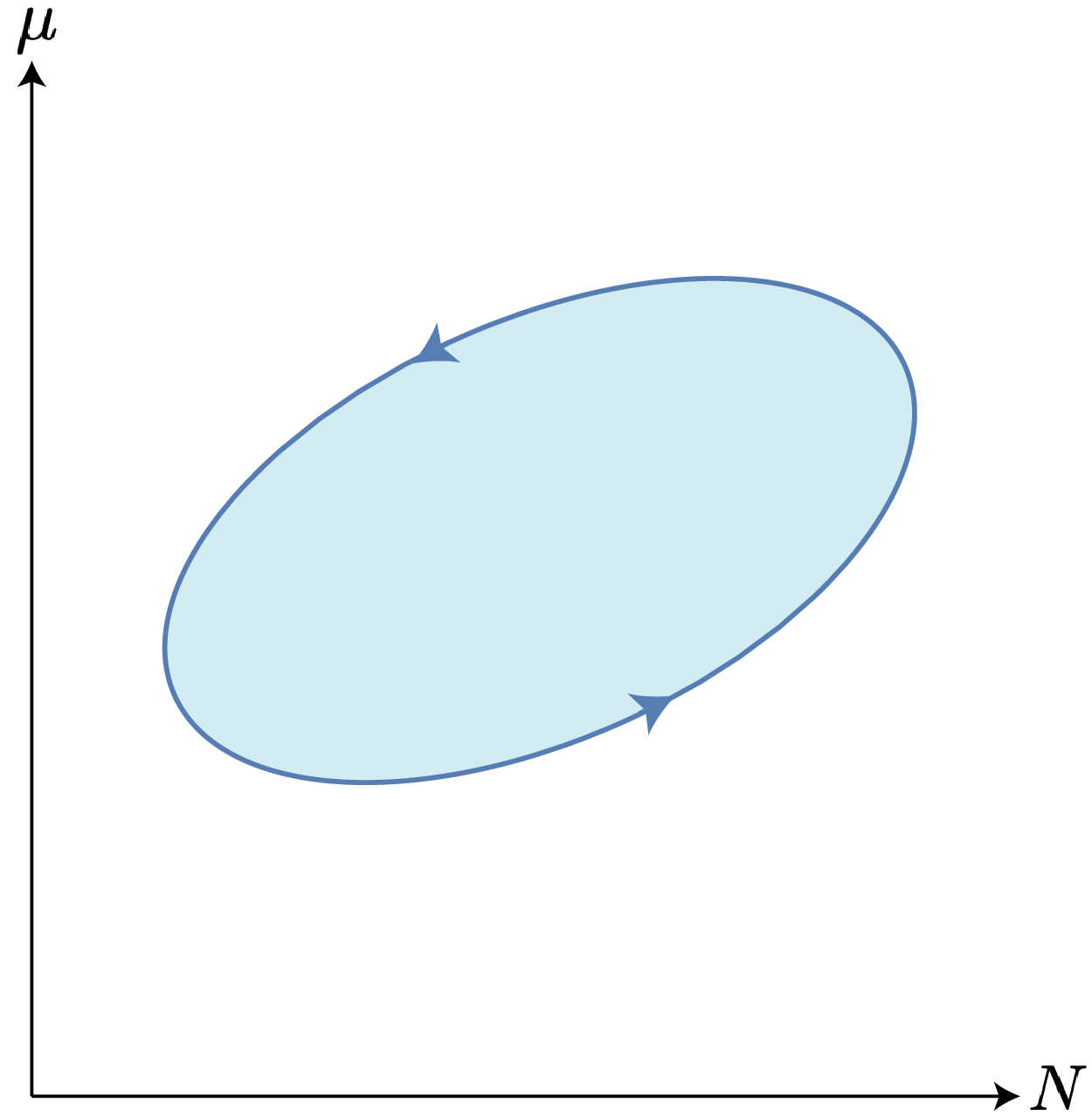}
\caption{\small Chemical-engine cycle represented in the $(\mu, N)$-plane.\label{fig:Nmu}}
\end{figure}

The second law of thermodynamics is satisfied because heat is dissipated into the environment  \hbox{($\delta {\cal Q} < 0$)} at constant temperature $T$, so that \hbox{$\dot S = - \dot{\cal Q} / T > 0$}.  Integrating \Eq{eq:1law} over a complete cycle of the system's thermodynamic state we obtain that
\be
W_0 \equiv \oint \delta W < \oint  V_d(Q,X) \, d{Q} =  \oint  \mu \, d N ,
\la{eq:cycle}
\ee
where $W_0$ is the useful work generated by one limit cycle of the self-oscillation.  The right-hand side of \Eq{eq:cycle} is the area contained within a closed trajectory on the $(\mu, N)$-plane, as shown in \Fig{fig:Nmu}.  This curve is the {\it thermodynamic cycle} of the half-cell considered as a chemical engine.

Equation \eqref{eq:cycle} implies that sustained work extraction ($W_0 > 0$) by an open system coupled to an external chemical disequilibrium requires the system to change its state in time in such a way that $\mu$ varies in phase with $N$.  This is a particular instance of what was called the generalized ``Rayleigh-Eddington criterion'' (after the physicists who clearly formulated this principle for heat engines) in Ref.\ \onlinecite{solar}.  A clear example of this principle for a microscopic chemical engine is provided by the electron shuttle.\cite{gorelik, wachtlerNJP, wachtlerPRA} The work $W_0$ extracted by the cycle can then be used to to {\it pump} electric charge $\bar Q$ {\it against} the time-averaged electric potential difference at the half-cell, which corresponds to the generation of an emf
\be
{\cal E} = \frac{W_0}{\bar Q} ~.
\ee

Note that the modulation of $\mu$ with respect to $N$, required by the Rayleigh-Eddington criterion, is possible only if the thermodynamic cycle (which in this case corresponds to the self-oscillation of the double layer) is slow compared to the time-scale of the chemical reactions that control the value of $\mu$ at each point within the cycle.  For instance, an automobile can run because it takes in pristine fuel at high chemical potential and expels burnt fuel at low chemical potential.\footnote{The internal combustion engine is usually conceptualized as a heat engine, with the air in the cylinder as working substance.  But it is also possible to consider it as a chemical engine, with the fuel-air mixture as working substance.  Of course, in the latter analysis, \Eq{eq:cycle} gives only a loose upper bound on the extracted work because of the large amount of heat (and therefore entropy) that the engine dumps into the environment when it expels the burnt fuel.}  Combustion must therefore proceed quickly compared to the period of the motion of engine's pistons.  If the time taken by the combustion were comparable to the period of the piston, $\mu$ would not vary effectively with $dN$ in \Eq{eq:cycle} and little or no net work could be done by the engine.

The self-oscillation of the double layer is also the {\it pumping cycle}, which converts mechanical into electrical work, as we shall discuss in more detail in \Sec{sec:pump}.  This pumping must therefore be slow with respect to the redox reactions from which the battery ultimately takes the energy to generate the electrical work.  This conceptual distinction between the fast chemical reaction and the slower pumping is, in our view, the key missing ingredient in all previous theoretical treatments of the battery.

%%%%%%%%%%
%%% CURRENT PUMPING
%%%%%%%%%%

\subsection{Current pumping}
\la{sec:pump}

Having described the extraction of mechanical work by the double layer's self-oscillation, we proceed to consider how that work can be used to pump an electrical current, thereby generating the battery's emf.  As shown in Ref.\ \onlinecite{LEC}, the instantaneous power that an irrotational electric field $\vv E (t, \vv r)$ delivers to a current density $\vv J (t, \vv r)$ contained in a volume $\cal V$ is
\be
P = \int_{\cal V} \vv E \cdot \vv J \, d^3 r = -  \int_{\cal V} \phi \frac{\partial \rho}{\partial t} \, d^3 r ,
\la{eq:power}
\ee
where $\phi(t, \vv r)$ is the potential (such that $\vv E = - \boldsymbol \nabla \phi$) and $\rho (t, \vv r)$ is the charge density (such that $\boldsymbol \nabla \cdot \vv J = - \partial \rho / \partial t$).

In an ordinary capacitor, electrical charge comes out ($\partial \rho / \partial t < 0$) of the positively-charged plate at potential $\phi_+$, while charge enters ($\partial \rho / \partial t > 0$) the negatively-charged plate at potential $\phi_-$.  The instantaneous power is therefore
\be
P = V \cdot I > 0 ,
\la{eq:IV}
\ee
where $V = \phi_+ - \phi_-$ and $I$ is the integral of $| \partial \rho / \partial t |$ over the complete volume of either plate.  A similar analysis applies to the discharging of the two double layers in the supercapacitor.

The case of the battery is more subtle, because it can generate $P > 0$ without appreciable accumulation or depletion of charge anywhere in the circuit.  Purcell argued that this apparent contradiction is resolved by the fact that the battery's operation involves chemical reactions, which must be described in terms of quantum mechanics.\cite{Purcell}  Indeed, the average flow of {\it matter} inside the battery can be described in terms of discharging {\it chemical} potentials,\cite{Newman} which represent an underlying quantum physics.  But the details of how this chemical energy is converted into electrical work, in a thermodynamically irreversible way and without contradicting the laws of classical electrodynamics, have not been adequately clarified in the literature.

It is well known that charges can be accelerated by {\it periodic} oscillations of $\phi$ and $\partial \rho / \partial t$, as is done in a modern particle accelerator.\cite{cyclotron}  As we shall see, the self-oscillation of the double layer described in \Sec{sec:double-layer} modulates $\partial \rho / \partial t$ in phase with $- \phi$, allowing net electrical work to be performed over a complete period of the oscillation.

When the double layer contracts (i.e., $\dot X < 0$), the Helmholtz layer is moving towards the electrode, and therefore along a direction of decreasing potential $\phi$.  The value of $\partial \rho / \partial t$ is negative in the region that the Helmholtz layer is moving out of (where the voltage is higher) and positive in the region that the Helmholtz layer is moving into (where the voltage is lower).  Equation \eqref{eq:power} therefore implies that $P > 0$ during the contraction phase.  This reflects the fact that the ion plane is moving along the internal electric field, and is therefore being accelerated by it.

On the other hand, when the double layer expands (i.e., $\dot X > 0$), the ion plane is moving into a region that is screened from the internal electric field.  Thus, the region where $\partial \rho / \partial t$ is negative is at a voltage only slightly lower than the region in which it is positive.  This allows the net work (i.e., the integral of $P$ in \Eq{eq:power} over a full period of the double layer's self-oscillation) to come out positive, which corresponds to a sustained pumping of the current.

\begin{figure} [t]
	\centering
	\includegraphics[width=0.49 \textwidth]{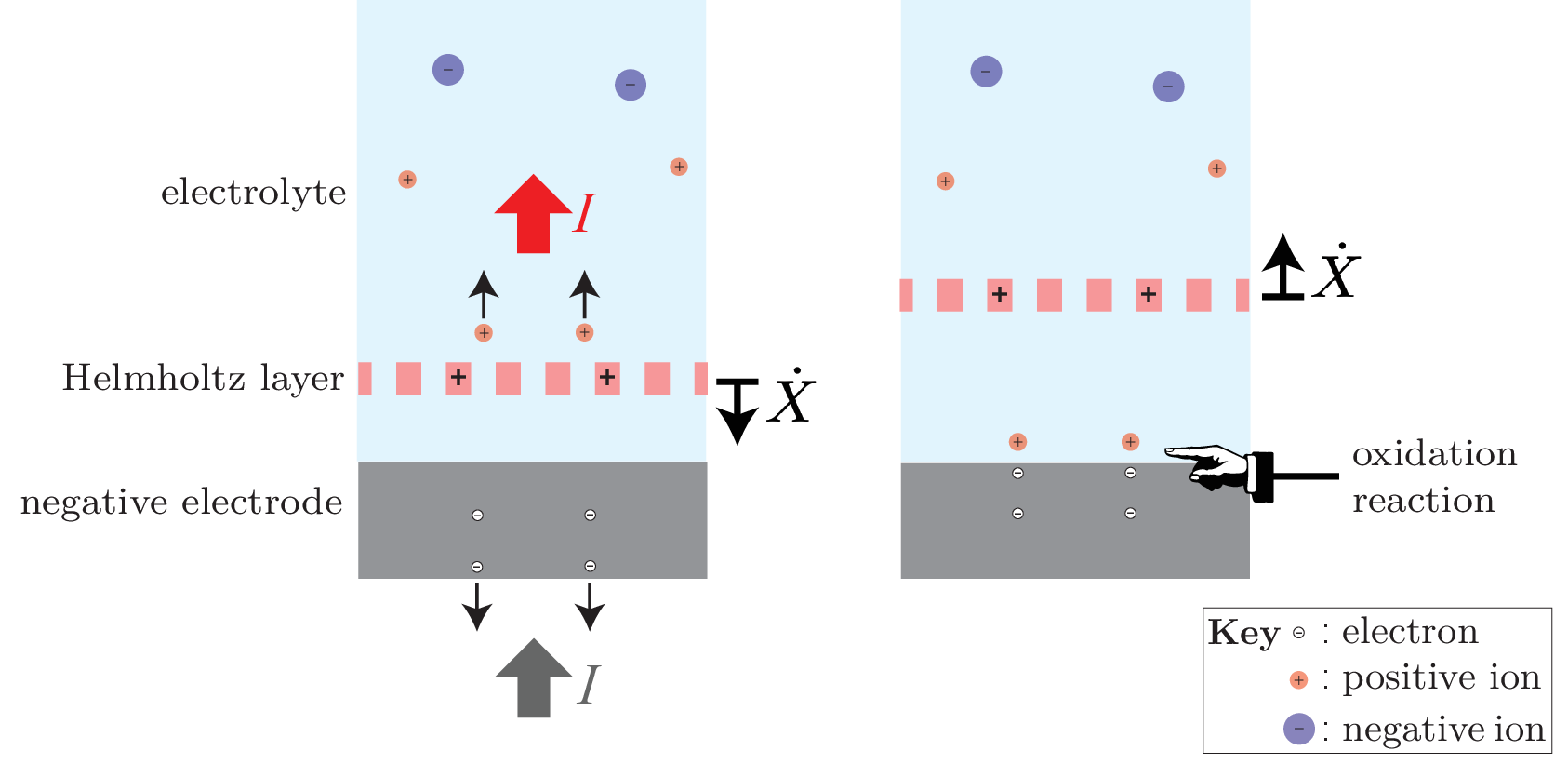}
\caption{\small On the left: during the compression of the double layer, positive ions are squeezed out of the double layer and into the bulk of the electrolyte, giving them ballistic rather than diffusive motion.  The Coulomb interaction then drives free electrons away from the interface and into the conducting terminal, maintaining charge balance.  On the right: no pumping of current occurs during the expansion of the double layer.  The ions and electrons pumped away during the compression phase are replenished by the oxidation reaction.\label{fig:pumping}}
\end{figure}

Intuitively, we may describe the pumping at the negatively charged terminal in the following terms:  During the contraction phase, the positive ions in the double layer are ``squeezed'', giving them a ballistic motion that allows some of them to pass through the Helmholtz layer and escape into the neutral bulk of the electrolyte, thus generating an active discharge current.  The current is rectified because the ions can pass through the Helmholtz layer but not penetrate the solid electrode.  The generation of a net current by this oscillation is similar to the operation of a ``valveless pulsejet'' engine.\cite{pulsejet}

The Coulomb interaction between the positive ions in the electrolyte and the electrons in the electrode enforces charge balance, causing an equivalent current to be drawn in from the external circuit and into the electrode during the squeezing phase.  This equilibration of charge happens very quickly compared to the double layer's self-oscillation, and it directs the flow of electrons in the external circuit.

During expansion, the interior of the double layer is replenished with positive ions and the electrode is replenished with electrons by the oxidation reaction that consumes the battery's ``chemical fuel'' in a thermodynamically irreversible way.  This sets up the double layer for the next pumping cycle.  The full process is illustrated in \Fig{fig:pumping}.  Note that if the positive ions are injected by the oxidation reaction when the double layer is expanded (so that $V_d$, and therefore the chemical potential $\mu$ for the ions, is high) and ballistically expelled when the double layer is contracted (so that $\mu = q V_d$ is low), \Eq{eq:cycle} implies that a net positive work can be done to pump the current $I$.

An analogous pumping may occur in the other half-cell.  Note, however, that pumping at only one of the half-cells may be sufficient to produce an emf.  Further details on the pumping dynamics probably depend on the specific battery configuration being considered.

From a microphysical point of view, this active transport of charges in an asymmetric potential subjected to a coherent, periodic modulation, is similar to the mathematical model of an artificial Brownian motor.\cite{Hanggi}  In the battery, it is the mechanical self-oscillation of the double layer that produces the periodic modulation of the potential for the ions.  Thermodynamically speaking, that modulation is the external work that drives the pumping and which is converted into the electrostatic energy of battery's charged terminals, minus some internal dissipation.

There is ample experimental evidence from ``sonoelectrochemistry'' that the application of ultrasound to an electrochemical double layer can significantly enhance the electrical current through that layer.\cite{Compton}  This effect has usually been explained as resulting from the ultrasonic driving causing the Nernst diffusion layer at the electrode-electrolyte interface to become thinner, but in our view it may more plausibly be interpreted as resulting from the pumping of current by the forced oscillation of the double layer, as described by \Eq{eq:power}.

%%%%%%%%%%
%%% COMPARISON WITH EXPERIMENT
%%%%%%%%%%

\section{Comparison with experiment}
\la{sec:experiments}

To estimate the oscillation frequency $\Omega_0$ we use the data from Ref.~\onlinecite{Favaro2016}:
\bea
X_0 &\sim& 10 ~\text{--}~ 30 ~\hbox{nm} , \nl
V_d(Q_0, X_0) &=& 150 ~\hbox{mV} .
\la{eq:data}
\eea
Putting $\epsilon = 80$ and $\rho =10^3$ kg/m$^3$ into \Eq{eq:Omega} gives $\Omega_0 \simeq 0.1 ~\text{--}~ 1$ GHz.  This is much faster than the self-oscillations that have been experimentally observed so far, in the range of Hz to kHz.  However, we have reason to believe that the double-layer of the battery half-cell can indeed oscillate coherently at this high frequency, as required by our model of charge pumping.  It has been reported recently that operation of Li-ion batteries responds strongly to an external acoustic driving at 0.1 GHz.\cite{acoustic} The present ability of experimenters to manipulate the electrode-electrolyte interface in Li-ion batteries using sound at these high frequencies also suggests that it might be possible to detect the double layer's self-oscillation directly, by looking for the acoustic signal that it produces.

The low frequency self-oscillations that have been reported experimentally must therefore be of a somewhat different nature to the mechanism described in \Sec{sec:double-layer}.  A useful analogy is to the relationship between the fast dynamics that gives hydrodynamic pumping and the slow, parasitic self-oscillations that can affect the performance of such pumps.  Those parasitic self-oscillations get their power from the fast pumping mechanism, but they have a different dynamics from that of the pumping itself.\cite{hydro-so}

One possible source of the slow electrochemical self-oscillations are mechanical surface waves moving along the Helmholtz plane of the double layer.  These may produce modulations of the output voltage that are superimposed on the fast ``pumping oscillations''.  Determining whether this is, in fact, the mechanism of some of the electrochemical self-oscillations that have been observed will require detailed numerical simulation of a model of the double-layer that incorporates a mechanical elasticity for the deformation of the Helmholtz plane away from its flat configuration, as well as spatial inhomogeneity of the charging process described by \Eq{eq:kinQ}.\cite{surface-waves}

Alternatively, the very slow oscillations (with frequencies \hbox{$1 ~\text{--}~ 10^{-4} $ Hz}) could be of purely chemical nature. To illustrate their possible dynamics we can replace \Eq{eq:kinQ} by a more complicated kinetic equation involving molecules $Q$ and $R$ with varying concentrations and molecules $A$ and $B$ provided by the chemical baths with fixed concentrations.  Consider the following set of irreversible chemical reactions containing two autocatalytic reactions:
\be
Q + R \rightarrow 2Q , \quad Q \rightarrow  A , \quad R + B \rightarrow 2 R ,
\la{eq:reaction}
\ee
with the rates $k_1 , k_2 , k_3$. The corresponding kinetic equations 
\bea
\dot Q &=& - k_2  Q  + k_1 Q R , \nl
\dot R &=& (k_3 B) R  -  k_1  Q R ,
\la{eq:LV}
\eea
implement the well-known Lotka-Volterra dynamical system, leading to self-oscillating concentrations.\cite{LotkaVolterra}

In this case, slow chemical oscillations are superimposed on the fast self-oscillations of the ``piston'' described in \Sec{sec:LEC}.  Note that, as the battery's chemical fuel starts to run out, the concentration of $B$ decreases and the stationary point \hbox{$(R_0 = k_2 / k_1 , Q_0 = k_3 B / k_1)$} of the kinetic \Eq{eq:LV} is shifted.  The slow chemical oscillations of $Q, R$ may therefore grow, which would be consistent with the behavior reported in Ref.~\onlinecite{Li}.  To obtain useful work from such a chemical self-oscillation some coupling to a mechanical degree of freedom (analogous to the $X$ of the LEC model) is needed.\cite{Isakova}

%%%%%%%%%%
%%% DISCUSSION
%%%%%%%%%%

\section{Discussion}
\la{sec:discussion}

We have proposed a dynamic model of the electrode-electrolyte interface that can explain the pumping that gives rise to the battery's emf.  This required us to move beyond the electrostatic description on which most of the literature in both condensed-matter physics and electrochemistry is based.  For this we have considered variations in time of the mechanical state and charge distribution of the double layer.  In thermodynamic equilibrium, the fluctuations of the double layer would quickly average out, yielding no macroscopic current.  But we have shown that an underlying chemical disequilibrium can, in the presence of a positive feedback between the mechanical deformation and the charging of the double layer, cause a coherent self-oscillation that can pump a current within the cell, charging the terminals and sustaining the current in the external circuit.  The relevant feedback mechanism is introduced by the dependence of the rates $r_\pm$ on $X$ and $Q$ in \Eq{eq:kinQ}.

The general idea that we have advanced here, that the battery's emf is generated by a rapid oscillatory dynamics at the electrode-electrolyte interface of the half-cell, was already suggested by Sir Humphry Davy (the founder of electrochemistry) in 1812:
\begin{quote}
It is very probable that [...] the action of the [solvents in the electrolyte] exposes continually new surfaces of metal; and the electrical equilibrium may be conceived in consequence, to be alternately destroyed and restored, the changes taking place in imperceptible portions of time.\cite{Davy}
\end{quote}
Although he lacked the conceptual tools to formalize this insight, which was therefore later lost, Davy understood that the action of the battery was at odds with a description in terms of time-independent potentials, as well as with a simple relaxation to an equilibrium state.

The very use of the term {\it electromotive force} (as distinct from the potential difference or voltage) points, in the context of the battery, towards an off-equilibrium, dynamical process that irreversibly converts chemical energy into electrical work, equal to the total charge separated across the two terminals times the potential difference between those terminals.  In classical electrodynamics, emf is often equated to circulation of the electrical field, but this is not relevant to the battery, in which no macroscopic and time-varying magnetic field is present (as would be required by the Faraday-Maxwell law to give electric circulation).\cite{Purcell, LEC}  

To date, no microphysical, quantitative description of the generation of the battery's emf has been worked out.  We believe that this theoretical blind spot arises from several factors.  Firstly, the emf cannot be directly measured, monitored, or detected in a discharging battery. Experimental studies of batteries rely on discharge curves that relate the potential between the two terminals to the charge over time. Secondly, the emf ---by its nature--- is not an electrostatic phenomenon, and therefore lies beyond the usual theoretical framework of electrochemistry.  That framework has proved powerful by making many useful and accurate predictions, but it neglects the dynamical nature of the operation of electrochemical cells.  As the historian and philosopher of science Hasok Chang has noted correctly, ``for anyone wanting a rather mechanical or causal story about how free electrons start getting produced and get moved about, the modern textbook theory is a difficult thing to apply.''\cite{Chang1}

This situation is not unusual in physics and chemistry. Classical thermodynamics can be used to predict the efficiency of a steam engine, with no knowledge of the mechanical or dynamical properties of the moving parts. However, as we have argued here, to fully understand the battery's emf we must consider such off-equilibrium dynamics.  Our description of the battery's emf is based on an active, non-conservative force that pumps the current in the circuit.  This leads to an account that requires conceptual tools not ordinarily taught to either chemistry or physics students, whose training in the dynamics of non-conservative systems is usually limited to stochastic fluctuations and the associated dissipation.

Our own thinking in this subject has been influenced by recent developments in non-equilibrium thermodynamics, and especially by research in ``quantum thermodynamics'' seeking to understand the microphysics of work extraction by an open system coupled to an external disequilibrium.\cite{QT}  For a recent theoretical investigation in which the microphysics of the generation an emf by a triboelectric generator is considered in those terms, see Ref.~\onlinecite{tribo}.  In the present case of the battery, however, our treatment has been wholly classical.

The observations of slow self-oscillations in certain battery configurations\cite{Li, Goodenough} have clearly established that the electrochemical double layer at the electrode-electrolyte interface is a complex dynamical system, capable of generating an active cycle through the interplay of its chemical, electric, and mechanical properties.  Together with the theoretical considerations, based on electrodynamics and thermodynamics, detailed in this article, we believe that this makes a strong case that the generation of the emf by the battery must be understood as an active periodic process, like other instances of the pumping of a macroscopic current.

The recent report of the response of the Li-ion battery to an external acoustic driving with a frequency of 0.1 GHz\cite{acoustic} is consistent with our theory and, in our view, points towards the rapid double-layer dynamic that we have identified as the mechanism responsible for the emf.  If our theory is correct then there must be detectable signals of this high-frequency oscillation.  Observation of the $0.1 ~\text{--}~ 1$ GHz electromagnetic radiation, which is strongly absorbed by water, may be experimentally challenging.  Other possible signals include fast residual modulations of the battery's output voltage (much faster than those reported in Refs.~\onlinecite{Li, Goodenough}) and the battery's response when subject to ultrasound signals capable of resonantly driving the double layer's oscillation.  Moreover, as noted in \Sec{sec:experiments}, if the present model were extended to include mechanical elasticity of the capacitor plates and spatial inhomogeneity of the charging processes, quantitative predictions of slower self-oscillations (in a range of frequency more accessible experimentally) might result.

Our model is consistent with the established principles for the optimization of battery design, based on increasing the chemical potential difference at the anode and cathode, and increasing the capacitance of each half-cell.  However, our dynamical treatment introduces a new consideration, namely the mechanical properties of the double layer, to which very little consideration has been given in the past.

The theory that we have proposed here for battery half-cell as a dynamical engine might be generalizable to other chemically-active surfaces, including the catalytic systems considered in Refs.~\onlinecite{Ertl} and \onlinecite{Delmonde}.  In those cases the work extracted from the external disequilibrium is used to drive a non-spontaneous chemical reaction, rather than to pump a macroscopic electrical current. The fact that some kinds of catalytic reaction are associated with micro-mechanical self-oscillations of the catalyst is already well established experimentally.\cite{Ertl} \\

%%%%%%%%%%
%%% ACKNOWLEDGMENTS
%%%%%%%%%%

\begin{acknowledgments} We thank Luuk Wagenaar and Lotte Schaap for fruitful discussions and critical questions.  AJ also thanks Esteban Avenda\~no, Diego Gonz\'alez, Mavis Montero, and Roberto Urcuyo for educating him on electrochemical double layers.  RA was supported by the International Research Agendas Programme (IRAP) of the Foundation for Polish Science (FNP), with structural funds from the European Union (EU). DG-K was supported by the Gordon and Betty Moore Foundation as a Physics of Living Systems Fellow (grant no.\ GBMF45130).  AJ was supported by the Polish National Agency for Academic Exchange (NAWA)'s Ulam Programme (project no.\ PPN/ULM/2019/1/00284).  EvH was supported by the research programme ENW XS (grant no.\ OCENW.XS.040), financed by the Dutch Research Council (NWO). EvH and RA also gratefully acknowledge the support of the Freiburg Institute of Advanced Study (FRIAS)'s visitors' program during the first stages of this collaboration. \end{acknowledgments}

%%%%%%%%%%
%%% REFERENCES
%%%%%%%%%%

%%%%%%%%%%
%%% APPENDIX
%%%%%%%%%%

\onecolumngrid
\newpage

\appendix*
\section{Double-layer dynamics and self-oscillation}
\la{sec:appendix}

In this Appendix we provide the details of the mathematical treatment of the LEC model for the battery, whose results are used in \Sec{sec:conditions}.  It is convenient to introduce dimensionless variables $x,y$ to characterize the deviation of capacitor plates from their equilibrium position and charge separation, respectively, so that
\be
X = X_0 (1 + x) \quad \hbox{and} \quad  Q = Q_0 (1 + y) ,
\ee
where $X_0$ and $Q_0$ give zero force in \Eq{eq:f}.  We also parametrize the capacitance as
\be
C(X) =  C_0 \frac {X_0}{X}= \frac{C_0}{1 + x} .
\ee
Equation \eqref{eq:Newt} then takes the form
\be
\ddot x + \gamma \dot x =  -\Omega_0^2 \left[ (1 + y)^2 - \frac{1}{1+x} \right] ,
\la{ap:x-eom}
\ee
where
\be
\Omega_0^2 = \frac{Q_0^2}{2M X_0^2 C_0}
\ee
is the square of the angular frequency $\Omega_0$ for small oscillations about equilibrium ($x=0$, $y=0$).

The capacitance of the LEC at equilibrium can be expressed in terms of surface area $\cal A$, plate separation $X_0$, and electric permittivity $\epsilon\epsilon_0$ of the electrolyte:
\be
C_0 =  \frac{\epsilon\epsilon_0 {\cal A}}{X_0} .
\ee
This leads to a useful expression for the frequency $\Omega_0$ in terms of the potential drop $V_{\rm d} (X , Q) =  Q / C(X)$ and the effective density $\rho = M / ({\cal A} X_0)$:
\be
\Omega_0^2 = \frac{\epsilon\epsilon_0 V_{\rm d}^2 (X_0 , Q_0)}{2 \rho X_0^4} ,
\ee
which corresponds to \Eq{eq:Omega} in the main text.  The density $\rho$ should be of the same order of magnitude as the electrolyte density.

To determine the condition for the onset of self-oscillations (the ``Hopf bifurcation''), it is enough to consider linear perturbations about equilibrium.\cite{Strogatz}  The linearized version of \Eq{ap:x-eom}, valid for $|x|, |y| \ll 1$, is
\be
\ddot x =- \gamma \dot x   -\Omega_0^2 ( x + 2 y) .
\la{ap:lin1}
\ee
Inserting the lowest order expansions of the rates in \Eq{eq:rates},
\be
r_{\pm}(X,Q) \simeq r_{\pm} + [\partial_X r_{\pm}](X - X_0) +  [\partial_Q r_{\pm}](Q - Q_0) ,
\ee
and the equilibrium relation
\be
\frac{r_+}{r_-}=  \frac{Q_0}{q}
\ee
(in obvious shorthand notation) into the kinetic \Eq{eq:kinQ}, we arrive at the linearized equation
\be
\dot y = - \Gamma y  + b x ,
\la{ap:lin2}
\ee
with 
\be
\Gamma = r_- (X_0 , Q_0 ) \left\{ 1 - Q_0 \frac{\partial}{\partial Q} \left[ \ln \frac{ r_+ (X_0 , Q )}{ r_- (X_0 , Q )} \right]_{Q=Q_0} \right\}
\ee
and
\be
b = r_- (X_0, Q_0) X_0 \frac{\partial}{\partial X} \left[ \ln\frac {r_+(X , Q_0)}{r_-(X, Q_0)} \right]_{X = X_0} .
\la{ap:b}
\ee

For $r_\pm$ given by \Eq{eq:rates}, and using \Eq{eq:Gibbs}, we have
\be
\partial_X \ln \frac{r_+}{r_-} = \partial_X \ln k_+ - \partial_X \ln r_- = \partial_X \ln \frac{k_+}{k_-} + \partial_X \ln \frac{k_-}{r_-}
= - \frac{\partial_X \Delta G}{RT} + \frac{\partial_X k_-}{k_-} - \frac{\partial_X r_- }{r_-} .
\la{ap:dxln}
\ee
Equation \eqref{eq:Gibbs-stab} implies that $\partial_X \Delta G = 0$.  Thus we have
\be
\partial_X \ln \frac{r_+}{r_-} = \frac{r_- \partial_X k_- - k_- \partial_X r_-}{k_- r_-} = \frac{k_{\rm out} \, \partial_X k_-}{k_- r_-} = \frac{k_{\rm out} \, \partial_X \ln k_-}{r_-} .
\la{ap:rpm}
\ee
Inserting \Eq{ap:rpm} into \Eq{ap:b} we obtain
\be
b = k_{\rm out} \, X_0 \partial_X \ln k_- ,
\ee
which corresponds to \Eq{eq:b-cond} in the main text.

In order to find an approximate solution to the linearized Eqs.\ \eqref{ap:lin1} and \eqref{ap:lin2}, we assume that the damping rate $\gamma$ and the reaction rates $\kappa_{\pm} $ are small compared to $\Omega_0$. Then we can expect a solution of the form
\be
x(t) = e^{\mu t} \cos(\Omega_1 t) , \quad  y(t) = A e^{\mu t} \cos(\Omega_1 t + \phi)
\la{ap:xsol}
\ee
with $\Omega_1 \simeq \Omega_0$ and  $|\mu| \ll \Omega_0$.  Inserting \Eq{ap:xsol} into Eqs.\ \eqref{ap:lin1} and \eqref{ap:lin2}, and comparing the coefficients of the functions $e^{\mu t} \cos(\Omega_0 t)$ and $ e^{\mu t} \sin(\Omega_0 t)$, one obtains four equations in four unknowns: $\mu, \Omega_1, A$, and $\phi$.  One can simplify this system of equations by putting $\Omega_1 = \Omega_0$ and neglecting small contributions like $\mu/\Omega_0 , \gamma/\Omega_0$, etc. This procedure leads to the final approximate expression 
\be
\mu \simeq b - \frac{\gamma}{2}
\ee
and the corresponding approximate threshold for the emergence of self-oscillations:
\be
b > \frac \gamma 2 .
\ee

\end{document}